\documentclass[a4paper]{article}

\usepackage[english]{babel}
\usepackage[utf8x]{inputenc}
\usepackage[T1]{fontenc}

\usepackage{amsthm}

\usepackage{url}

\usepackage[a4paper,top=3cm,bottom=2cm,left=3cm,right=3cm,marginparwidth=1.75cm]{geometry}

\usepackage{amsmath}
\usepackage{graphicx}
\usepackage[colorinlistoftodos]{todonotes}
\usepackage[colorlinks=true,allcolors=black]{hyperref}

\newtheorem{dummy}{Dummy-Theorem}[section]
\newtheorem{definition}[dummy]{Definition}
\newtheorem{protocol}[dummy]{Protocol}

\title{A Predictable Incentive Mechanism for TrueBit}
\author{Julia Koch\\ \texttt{schiefkoerper@gmail.com}
\and
Christian Reitwießner\\ \texttt{chris@ethereum.org}}

\begin{document}
\maketitle

\begin{abstract}
\noindent
TrueBit~\cite{truebit} is a protocol that uses interactive verification to allow a resource-constrained
computation environment like a blockchain to perform much larger computations than usual
in a trusted way. As long as a single honest participant is present to verify the computation,
an invalid computation cannot get accepted.

\noindent
In TrueBit, the presence of such a verifier is incentivised by randomly injected forced errors.
Additionally, in order to counter sybil attacks, the reward for finding an error drops off exponentially
with the number of challengers.

\noindent
The main drawback of this mechanism is that it makes it very hard to predict whether participation will be
profitable or not.

\noindent
To even out the rewards, we propose to randomly select multiple solvers from a pool and
evenly share the fees among them, while still allowing outside challengers.
Furthermore, a proof of independent execution will make it harder to establish computation
pools which share computation results.
\end{abstract}

\section{Introduction}

Due to their consensus construction, it is very hard to inject invalid computations
into a blockchain. This is in part achieved by the fact that every computation is checked
by everyone else participating in the system. This of course creates a scalability problem
and thus the amount of computation per unit of time is limited.

TrueBit~\cite{truebit} is a mechanism to get around this problem: Only the result of
a computation is submitted to the blockchain. Parties interested in this computation can
(but not all parties involved in the blockchain have to) re-compute it and challenge the submitter in case
they obtain a different result.
The challenge leads to an interactive binary search over the steps in the computation
exposing a faulty step which is then checked by the blockchain itself.

Apart from timeouts, it is impossible for an incorrect result to withstand such a challenge.
Because of that, TrueBit has so-called ``unanimous consensus'', since it only takes a single
honest person to expose the error.

This means the main problem to solve is how to ensure that such an honest person actually
participates. If we assume a certain minimum percentage of honesty, it is sufficient to
ensure that a certain number of participants checked the computation,
but it is an open problem how to prove that you checked a computation,
especially in the case that it was correct.

Fully homomorphic encryption can be used so that both the input and the computation is encrypted
to a key unknown to the verifier. The verifier then performs the computation on encrypted data and submits the
solution. After the solution has been submitted, the private key is published and everyone can compare the
results. This method of course has the drawback that fully homomorphic encryption is not yet practical, but it can
also be attacked if the verifier somehow can get access to the private key prior to the computation.

TrueBit tackles the problem of finding out whether a computation was actually checked
by randomly injecting invalid computations and rewarding people if they detect it.
Because of that, every computational task has a certain probability to contain a so-called
``forced error'', which creates an incentive to verify it. The problem with this is the way the rewards
are paid out: In order to protect against sybil attacks, the total payout is proportional to  $2^{-k}$ in case
there were $k$ participants. The formula is the result of the requirement that your total
reward decreases if you create an additional sybil account to report from.

In this article, we propose two changes to the protocol proposed in TrueBit~\cite{truebit}:
\begin{enumerate}
\item Instead of a single solver, a certain number of multiple solvers are selected from a pool of
idle and bonded solvers. If there is no disagreement, the reward is split evenly.
Verification games are played if the solvers disagree,
but we alse allow outside challengers. Succeeding outside challengers are still rewarded with exponential dropoff.
\item Together with the solution, solvers are required to submit a so-called
``proof of independent execution''~\cite{indiex}: The root of a Merkle tree
of state snapshots at each step which also contain a private secret, plus
a Merkle proof for a random leaf. This functions as a protection against pooling
and sharing results and replaces the probabilistic forced error and jackpot mechanism.
\end{enumerate}

In a sense, the difference between the original Truebit protocol and our
version mirrors the difference between miners' behaviour in Bitcoin
vs.\ Ethereum: Miners in Bitcoin only need to include transactions and
should check their validity, but might be tempted not to do so, the
so-called verifier's dilemma~\cite{Luu_demystifyingincentives}, whereas
in Ethereum not checking is not an option since a miner has to include
a post state root and simply giving a random one would be detected very
easily. In the same spirit, simply not checking solutions might be tempting
for a verifier in the original version of Truebit, while in our version
each chosen solver has to submit something and cannot simply not react.

\section{Protocol}
We assume familiarity with the TrueBit protocol and general terms
and definitions used in computer science and the blockchain space.

There is always a minimum of two solvers chosen for every task; if a task giver wishes
to obtain a result with a higher level of security, they can opt to have a larger number
of solvers.

To make sure that the solvers do calculate the solution independently,
they have to deliver a personalized proof of independent execution~\cite{indiex}.

\begin{definition}
A \emph{solver} is a participant in the protocol with a bonded deposit $D$.\\
A solver is called \emph{available} if they are not currently participating
in solving a task or a verification game. An account is \emph{slashed}
by destroying its deposit.
\end{definition}

\begin{protocol}
We propose the following protocol, where $H$ is a hash function.

\emph{Step 1:} Task giver provides task $T$ (computable in $n$ steps) and desired
number of solvers $k \ge 2$

\emph{Step 2:} A set of $k$ solvers $S$ is chosen randomly from the available solvers.
In the following, ``solver'' only refers to any of these selected solvers.

\emph{Step 3:} Within a certain timeout, each solver, identified by their address $s$, selects
a private random string $p$ and computes
the solution to $T$. While computing the solution, each solver creates hashes of state
snapshots $s_1, \dots, s_n$ and a Merkle tree
with leaves $L_1, \dots, L_n$, $L_i = H(s_i, H(s, p))$ and root $r$.
The values $H(s, p, y), r, pr(r)$ are submitted to the blockchain, where $y$ is the solution and $pr(r)$ is a Merkle proof
to the leaf $L_{1 + r \% n}$. The smart contract verifies the Merkle proof and only stores
$H(s, p, y)$ and the leaf $L_{1 + r \% n}$.

\emph{Step 4:} The first challenge period begins: everybody can stake a deposit and reveal a solver's private
randomness $H(s, p)$ resulting in the solver being slashed if correct.

\emph{Step 5:} After the timeout and another timeout to avoid potential rollbacks,
each solver in $S$ submits their own $p, y$.

\emph{Step 6:} All solutions are compared. If they disagree, verification games are played
tournament-style between solvers with different solutions (randomly selecting one
if there are multiple with the same solution) until only one solution remains.
At the same time, another challenge period starts: anyone can stake a deposit and
initiate a verfication game by challenging the correctness of either
a solution or of the intermediate state stored in the submitted Merkle leaf.

\emph{Step 7:} If a solution remains that was not successfully challenged, it is
accepted after a centain time of inactivity. If all solvers are slashed,
the protocol restarts from step 2.

\paragraph{Time-out:}
If a solver does not act within a given period of time, they lose
their deposit and a new solver is chosen.

\paragraph{Rewards:}
Someone who can guess a solver's private randomness receives half of their deposit.
Each non-slashed solver receives $X$ from the task giver.
Each non-slashed outside challenger receives $\frac{k\cdot D}{2^n}$, where
$n$ is the total number of non-slashed outside challengers and $k$ the number of
participants slashed in verification games.

\paragraph{Penalties:}
At the end of a verification game, the losing party loses their deposit $D$.

Remaining funds are burned.

To sum up, the deposit of a solver or outside challenger is slashed in any of the following situations:
\begin{itemize}
\item someone can guess their private randomness before the end of the submission timeout (only solvers)
\item the Merkle proof is invalid (only solvers)
\item they lose a verification game
\item they do not send a required message before a timeout occurs
\end{itemize}
\end{protocol}

We admit the possibility of an ``outside'' challenger
as a line of defense against cartels: If all solvers chosen for a task
conspire to deliver an incorrect solution, the dominant strategy for every single
participant in this conspiracy is to challenge the incorrect solution as an
outside challenger (with an account that the other conspirators do not know to
belong to her) to obtain (at least a percentage of) the slashed deposits of the
conspirators. Forming a cartel can thus never be a Nash equilibrium. 

The Merkle proof is given together with the root to help
other participants to check the correctness of the Merkle root after the private
randomness has been revealed:
If there is no Merkle proof, the verifier has to read the full state snapshots from disk and
create the Merkle root. If the Merkle proof is given right away, only a single
state snapshot has to be retrieved from disk.
Note that the Fiat-Shamir heuristic~\cite{FiatShamir} is used to determine
which Merkle proof to provide. This ensures that the full Merkle tree has been
created and the leaf is not fabricated.

\section{Security}

In the following, we analyse the security of the protocol.
We distinguish mere deviations from the protocol which try to reduce personal costs
and actual attacks which aim for an invalid solution being accepted.

\subsection{Possible Deviations from the Protocol}

\paragraph{Sharing Solutions}
If multiple solvers collaborate to avoid parallel execution, it is not enough
to just share the solution, because each solver has to provide an individual
Merkle proof. We conjecture that there is no way to generate a correct
Merkle proof without either knowing the private randomness or the full
sequence of state snapshots. Both pieces of information have to be on
the same computer to form a correct Mekle proof. Since we assume that
transmitting the state snapshots is costly compared to computing them,
the only reasonable way is to transmit the private randomness (or
directly generate it at the executing computer), in which case
the executing party can slash the benefiting solver.
Furthermore, slashing can be done pseudonymously, which is important in the
case where more than two solvers collaborate.

\paragraph{Pooling Execution}
One of the main problems in TrueBit is the impossibility to disincentivize participants from forming pools.

As long as they form a ``lazy cartel'', i.e.\ a pool that agrees on sharing solutions (in our version)/not challenging each others' solutions (in the original version), but still every member of the pool conducts their calculations honestly, this is per se not a problem for the trustworthiness of the system (though it skews the effort/reward ratio in favour of those who pool and might thus demotivate other participants over the course of time, and a pool that controls a certain percentage of all participants could very well threaten the integrity of the system). The risk is obviously that one member hands in a wrong solution and the others copy this solution/do not challenge it. 

Note that handing in a wrong solution will be punished in our version, while simply not challenging is not, so the ``copying solver'' bears a slightly higher risk; but the fundamental problem that such a wrong solution simply might not be detected remains. 

We further believe that our version removes one of the main motivations behind forming pools, though: Prospect Theory~\cite{KahnemanTversky} shows that human beings usually pick a strategy that delivers a predictable reward reliably over one that delivers even a greater expected value but bears a higher risk. Since the default strategy in our version of the protocol is of the former type, the need to ``spread the risk'' via some kind of ``insurance pools'' is removed. 

Moreover, even in the case where solvers take the risk to be slashed by collaborating,
it is not profitable to share the execution of tasks unless
the number of collaborating solvers is a large percentage of all solvers:
The number of computations can only be reduced if
those accounts are selected for the same tasks at the same time.
In addition to that, someone has to
create and maintain specialized software and pay for computing capacity for the pool.

\subsection{Possible Attacks}

In the original version of TrueBit, a wrong solution given by a solver
will be detected if and only if at least one honest verifier is checking this solution. 

In our version, a wrong solution given by a solver will be detected if another
(honest) solver for the same task gives the correct (differing) solution, or
if an outside challenger challenges the solution (in the case of no honest solver
within those chosen for the task). 

We calculate the security for our protocol: Assume that an attacker controls $q$
entities of the $n$ active participants in the solver pool. If the task-giver
chooses to have $k\geq 2$ solvers for their task, the attacker can get away with
a wrong solution if and only if they get randomly chosen for all $k$ roles. The probability for this is
\[\frac{\binom{q}{k}}{\binom{n}{k}} = \frac{q\cdot(q-1)\cdot\dots\cdot(q-k+1)}{n\cdot(n-1)\cdot\dots\cdot(n-k+1)}.\]

If an attacker controls at most $\frac1r$th of all solver accounts and a task giver chooses to have $k$ solvers, the attacker can thus run a successful attack with a probability of
\[\frac{\binom{q}{k}}{\binom{rq}{k}} = \frac{q\cdot(q-1)\cdot\dots\cdot(q-k+1)}{rq\cdot(rq-1)\cdot\dots\cdot(rq-k+1)}\leq
\frac{q\cdot(q-1)\cdot\dots\cdot(q-k+1)}{rq\cdot(rq-r)\cdot\dots\cdot(rq-r(k-1))} =\frac{1}{r^k}
.\]

If we take the same assumption as in the original paper, i.e., the attacker controls at most $\frac16$ of all solver accounts, this hence yields a probability of a successful attack of
\[\frac{\binom{q}{2}}{\binom{6q}{2}} = \frac{q\cdot(q-1)}{6q\cdot(6q-1)}\leq \frac{1}{36}\approx 2.8\%\]
in the basic case of two solvers, and a probability of less than $0.46\%$ in the case of three solvers. Note that this is at least as secure as the original protocol, especially if taking into account that the number of active verifiers for a task will decrease if the mechanism described in the security patch in \cite{truebit}, A1 ``Incorrect  secondary  solution'' is implemented.

We would furthermore like to stress that our protocol does not need any assumptions on (global) tax rates, but each task giver can choose their preferred level of security independently for every single task by adjusting the number of solvers.

\paragraph{Availability Engineering.}
As explained above, it is necessary to have control over all selected solvers
in order to get an invalid solution accepted. Since selected solvers need
to be available, one strategy would be to tweak the availability of solvers.

A solver can transition from available to unavailable either by being selected
or by acting as an outside challenger. Unavailable solvers can get available
by the task or verification game under consideration ending.

The unavailability of solvers not under control can be extended
by challenging them, but this will result in a deposit being slashed
(apart from the unlikely event that they actually are in error).

Another strategy would be to issue bogus tasks and hope that other solvers
get selected and only solvers under control remain, but the probability for
this happening is similar to just being selected in the original pool.

Making solvers under control unavailable such that they get available again
at a certain time could be another strategy, but the only way a solver can
get unavailable in a predictable way is by acting
as an outside challenger, which will result in being slashed unless there
is an actual error, which is unlikely. Producing an error on purpose also
likely results in a deposit being slashed.

\paragraph{Delays.}
An attacker can delay a task being resolved by continously acting as an outside challenger.
Since a challenger can lose the verification game even if the challenged solution is wrong,
this might be a strategy to ``bore'' other potential challengers. An unsuccessful challenge
always leads to slashing, though, and thus is way too expensive.

\subsection{Tradeoffs}
When comparing our proposed protocol with the original Truebit version, we see the following tradeoffs:

\begin{itemize}
\item Our version removes the need for forced errors, thus making jackpots, taxes, and, finally, even having a native Truebit token obsolete. It is thus a simpler, more intuitive protocol. 
\item Our version allows each task giver to choose their preferred level of security (by stating the number of solvers they wish) and pay accordingly. In the original version, the level is the same for each and any task and is determined by the tax rate. 
\item Psychology research suggests that our version would make it less attractive for participants to form pools: We conjecture that the highly unpredictable nature of payouts in the original Truebit protocol makes ``insurance pools'' that distribute the jackpot payouts (and especially the risk of not getting one) more evenly on a larger number of participants. Since our version yields lower, but steady income for solvers, this need is removed. 
\item If pools do form nevertheless, they might be more dangerous in our version since it does not usually encourage verifiers to double-check the solvers' solutions. If a pool gets ``openly'' dishonest (i.e., in the sense of all pool members knowing that incorrect solutions will be given), our provision for outside challengers makes betraying the pool the dominant strategy, but this is not neccesarily  a measure of security against ``lazy'' pools with a pool administrator gone rogue. 
\item Our version makes it more costly for a solver to hand in a solution (and thus also more costly for a task giver) since they need to run not only the calculation to solve the task itself, but also those necessary for the proof of independent execution. Note, though, that we do not need taxes, and that e.g.\ the security calculation in \cite{truebit}, p.~31, assumes that the tax rate supports 6 verifiers per task -- which would correspond to 7 solvers in our version of the protocol. To reach a similar level of security, we only need 3 solvers as shown on p.~5. Furthermore, the subdivision into elementary steps and they way the Merkle tree is defined leaves open several parameters which need to be tuned in practice and can heavily influence the computational resources needed.
\end{itemize}

The psychological/economic nature of these questions makes it hard to impossible to prove superiority of one protocol over another; we think that only empirical observation can show which one should be chosen finally.

\bibliographystyle{alpha}
\bibliography{bibliography}

\begin{thebibliography}{LTKS15}

\bibitem[Dra18]{indiex}
Justin Drake.
\newblock Proof of independent execution.
\newblock \url{https://ethresear.ch/t/proof-of-independent-execution/1988},
  2018.

\bibitem[FS87]{FiatShamir}
Amos Fiat and Adi Shamir.
\newblock How to prove yourself: Practical solutions to identification and
  signature problems.
\newblock In {\em Proceedings on Advances in cryptology---CRYPTO '86}, pages
  186--194. Springer-Verlag, 1987.

\bibitem[KT79]{KahnemanTversky}
Daniel Kahneman and Amos Tversky.
\newblock Prospect theory: An analysis of decision under risk.
\newblock {\em Econometrica}, 47(2):263--291, 1979.

\bibitem[LTKS15]{Luu_demystifyingincentives}
Loi Luu, Jason Teutsch, Raghav Kulkarni, and Prateek Saxena.
\newblock Demystifying incentives in the consensus computer.
\newblock In {\em Proceedings of the 22nd ACM SIGSAC Conference on Computer and
  Communications Security}, CCS '15, pages 706--719. ACM, 2015.

\bibitem[TR17]{truebit}
Jason Teutsch and Christian Reitwiessner.
\newblock A scalable verification solution for blockchains.
\newblock \url{https://people.cs.uchicago.edu/~teutsch/papers/truebit.pdf},
  2017.

\end{thebibliography}

\end{document}